\documentclass[journal]{IEEEtran}
\usepackage{amsmath,amsfonts,amssymb}
\usepackage{algorithmic}
\usepackage{array}
\usepackage[caption=false,font=normalsize,labelfont=sf,textfont=sf]{subfig}
\usepackage{textcomp}
\usepackage{stfloats}
\usepackage{url}
\usepackage{verbatim}
\usepackage{graphicx}
\def\BibTeX{{\rm B\kern-.05em{\sc i\kern-.025em b}\kern-.08em
    T\kern-.1667em\lower.7ex\hbox{E}\kern-.125emX}}
\usepackage{balance}

\usepackage{booktabs}
\usepackage{bm}
\usepackage{siunitx}
\usepackage{multirow}
\usepackage{colortbl}
\usepackage{nicematrix}
\usepackage{diagbox}

\def\C{\mathbb{C}}
\def\E{\mathbb{E}}

\def\L{\mathcal{L}}
\def\N{\mathcal{N}}

\def\thineq{\hspace{-.1em}=\hspace{-.1em}}

\newcommand{\norm}[1]{\left\lVert #1 \right\rVert}
\newcommand{\abs}[1]{{\left\lvert #1 \right\rvert}}

\def\STFT{\mathrm{STFT}}

\DeclareMathOperator*{\argmax}{argmax} % \, for the space
\DeclareMathOperator*{\argmin}{argmin}

\usepackage{pifont}% http://ctan.org/pkg/pifont

\newcommand{\DRR}{\mathrm{DRR}}
\newcommand{\rtsixty}{\mathrm{RT_{60}}}
\newcommand{\EDC}{\mathrm{EDC}}
\usepackage{xcolor}
\usepackage[normalem]{ulem}

\newcommand{\mf}[1]{{\textcolor{black}{#1}}}
\newcommand{\mfcmt}[1]{\textcolor{blue}{[MF: #1]}}

\newcommand{\lreplace}[2]{#2}

\usepackage[numbers,sort&compress]{natbib}
\makeatletter
\def\ps@IEEEtitlepagestyle{%
  \def\@oddfoot{\mycopyrightnotice}%
  \def\@evenfoot{}%
}
\def\mycopyrightnotice{%
  \begin{minipage}{\textwidth}
  \centering
  \footnotesize \copyright 2026 IEEE. Personal use of this material is permitted. Permission from IEEE must be obtained for all other uses, in any current or future media, including reprinting/republishing this material for advertising or promotional purposes, creating new collective works, for resale or redistribution to servers or lists, or reuse of any copyrighted component of this work in other works.
  \end{minipage}
}
\makeatother

\begin{document}
\title{
U-DREAM: Unsupervised Dereverberation\\guided by a Reverberation Model
}
\author{Louis Bahrman, Marius Rodrigues, Mathieu Fontaine, Gaël Richard
\thanks{L. Bahrman, M. Rodrigues, M. Fontaine, and G. Richard are with the Laboratoire de Traitement et Communication de l’Information (LTCI), Télécom Paris, Institut Polytechnique de Paris, 91120 Palaiseau, France. (e-mail: \{louis.bahrman, marius.rodrigues, mathieu.fontaine, gael.richard\}@telecom-paris.fr)
This work was funded by the European Union (ERC, HI-Audio, 101052978). Views and opinions expressed are however those of the authors only and do not necessarily reflect those of the European Union or the European Research Council. Neither the European Union nor the granting authority can be held responsible for them.
}
}

\maketitle
\begin{abstract}

This paper explores the outcome of training state-of-the-art dereverberation models with supervision settings ranging from weakly-supervised to virtually unsupervised, relying solely on reverberant signals and an acoustic model for training. Most of the existing deep learning approaches typically require paired dry and reverberant data, which are difficult to obtain in practice. 
We develop instead a sequential learning strategy motivated by a maximum-likelihood formulation of the dereverberation problem, wherein acoustic parameters and dry signals are estimated from reverberant inputs using deep neural networks, guided by a reverberation matching loss.
Our most data-efficient variant requires only 100 reverberation-parameter-labeled samples to outperform an unsupervised baseline, demonstrating the effectiveness and practicality of the proposed method in low-resource scenarios.
\end{abstract}


\section{Introduction}

\IEEEPARstart{A}{coustic} waves propagation in enclosed environments is significantly influenced by reflections and diffractions from surrounding surfaces and objects. These interactions alter the original waveform and result in reverberation, which can be modeled as a superposition of delayed and attenuated versions of the source signal.
Reverberation has long been recognized as a critical factor affecting speech intelligibility~\cite{boltTheorySpeechMasking1949}, and its detrimental effects on audio clarity have motivated decades of research.

The task of reverberation suppression, commonly referred to as dereverberation, has received renewed attention in recent years due to its relevance in a wide range of audio processing applications. Effective dereverberation is essential in enhancing the performance of hearing aids~\cite{hawkinsSignaltoNoiseRatioAdvantage1984}, improving communication quality in hands-free~\cite{habetsJointDereverberationResidual2008} \lreplace{and hand-held~\cite{jeubWeNeedDereverberation2010}}{} telephony, and enabling robust Automatic Speech Recognition (ASR) in human-machine interaction scenarios~\cite{yoshiokaMakingMachinesUnderstand2012}. It also serves as a key preprocessing step in general-purpose speech enhancement frameworks~\cite{vincentAudioSourceSeparation2018}.

Beyond suppression, reverberation itself plays a constructive role in audio production, particularly in simulating desired acoustic characteristics in post-processing. Reverberation conversion, or acoustic transfer, aims to transform a given recording, possibly containing unknown or undesired room effects, into a version consistent with a target acoustic environment. This can be achieved either through sequential dereverberation followed by reverberation synthesis~\cite{imDiffRENTDiffusionModel2024}, or through end-to-end approaches that perform direct transformation between reverberant conditions~\cite{sadokAnCoGenAnalysisControl2025}. In both cases, accurate modeling of room acoustics remains essential for achieving perceptually plausible results.

The monaural dereverberation task constitutes a linear blind inverse problem, which is inherently ill-posed, as the solution  is not uniquely determined. To address such ill-posedness, it is common to incorporate prior knowledge about the source or the degradation process.
Recently, hybrid approaches that combine model-based formulations with data-driven learning have gained traction in the audio processing community~\cite{gannotSpecialIssueModelBased2024}.

Despite their empirical success, data-driven methods, particularly those based on deep neural networks (DNNs), typically require large volumes of supervised training data in the form of paired dry and reverberant signals. These dry signals must be recorded in anechoic conditions, rendering data collection expensive and impractical. Furthermore, supervised systems often exhibit limited generalization to unseen reverberant conditions, reducing their robustness in real-world scenarios. Even unsupervised methods that learn dry speech priors without requiring paired data remain limited by the availability of dry recordings.

To overcome these constraints, we introduce a monaural dereverberation framework 
that operates in a virtually unsupervised manner, relying solely on reverberant signals for training. 
Building upon the hybrid dereverberation model proposed in \cite{bahrmanHybridModelWeaklySupervised2025}, we introduce an enhanced virtually unsupervised learning strategy that resolves the key limitations of the original approach:
\begin{itemize}
    \item  Unlike the earlier method which required a large dataset of acoustic parameters for training,
we demonstrate strong performance using only a small subset derived from 100 Room Impulse Responses.
\item We derive a novel maximum likelihood (ML) formulation for dereverberation guided by a parametric reverberation model. While ML-based methods have been proposed for multichannel settings~\cite{nakataniSpeechDereverberationBased2010,sekiguchiAutoregressiveMovingAverage2022},
this is, to the best of our knowledge, the first application of such a formulation to monaural dereverberation with explicit use of acoustic parameters.
\item Whereas the previous work was validated only on synthetic signals, our method is evaluated under realistic acoustic conditions, where it maintains competitive dereverberation performance despite the limited data regime.
\end{itemize}
To promote reproducibility and further research, we release our codebase, pre-trained models, audio and examples along with additional results\footnote{\url{louis-bahrman.github.io/UDREAM/}}.
% Our approach integrates an explicit acoustic model into the dereverberation process and offers two main contributions:

% 

The remainder of this paper is structured as follows. Section~\ref{sec:related_works} surveys related works on dry source and reverberation modeling for dereverberation. Section~\ref{sec:notations} introduces the reverberation model employed in this work. Section~\ref{sec:method} presents our methodological framework in detail. Experimental setup and evaluation results are reported in Sections~\ref{sec:experimental_setup} and~\ref{sec:results_and_discussion}, respectively. Finally, Section~\ref{sec:conclusion} concludes the paper.

\section{Related works}\label{sec:related_works}
In this section, we review existing deep-learning-based dereverberation approaches leveraging reverberation models. 
Existing monaural dereverberation methods can be categorized
based on their learning strategy, and reverberation model assumptions.
Speech models learning strategies are usually linked to the requirements of the training data: supervised models using paired-data or dry signals and unsupervised models using wet data. 
Reverberation model assumptions span from explicit Room Impulse Responses (RIRs), scalar and interpretable reverberation parameters (e.g., $\rtsixty$, $\DRR$), to autoregressive models (AR) and even methods that forgo any reverberation modeling.

The most prevalent family of dereverberation methods in recent literature leverages supervised learning using both wet and dry data with no explicit reverberation model. They often rely %via 
on fully data-driven models representing wet-to-dry mappings using
for instance phase-agnostic sub-band processing~\cite{weningerSpeechEnhancementLSTM2015}, phase-aware cross-band processing~\cite{haoFullsubnetFullBandSubBand2021} or state-of-the art attention-based architectures~\cite{saijoTFLocoformerTransformerLocal2024}.

A second class of approaches rely on the combination of an autoregressive model and a speech prior and includes the classical Weighted Prediction Error (WPE) method~\cite{nakataniSpeechDereverberationBased2010} and its deep-learning-based extensions~\cite{kinoshitaNeuralNetworkBasedSpectrum2017, yangIntegratingDataPriors2024}. 
Such extensions can be trained in an unsupervised manner from wet signals~\cite{petkovUnsupervisedLearningApproach2019} in order to reduce the computational cost of WPE.
While WPE-based methods leverage the AR model using backward linear prediction, recent approaches such as Forward Convolutive Prediction (FCP) extend this paradigm by modeling reverberation as a forward convolutional process, enabling more accurate and learnable representations aligned with neural architectures. %changed by GR
Such approaches are used in supervised~\cite{wangConvolutivePredictionMonaural2021} and unsupervised settings~\cite{wangUSDnetUnsupervisedSpeech2024}.
²
A third category of methods gathers 
supervised discriminative approaches, 
trained using not only pairs of dry and wet signals, but also more accurate AR models defined using reverberation scalar parameters or even the full RIR itself at training time. 
These approaches are used to inform a dereverberation model using acoustic information. For instance, the reverberation time is used
to increase performance by adjusting the Short-Time Fourier Transform ($\STFT$) parameters in~\cite{wuReverberationTimeAwareApproachSpeech2017}, a dereverberation model is jointly optimized with a reverberation time estimator to steer an attention module in~\cite{wangTeCANetTemporalContextualAttention2021}, or the dereverberation model is informed using the reverberation time estimated from a pre-trained DNN in~\cite{liCompositeT60Regression2023}.
Another advantage of leveraging reverberation parameters at training is that they can be modified at inference, enabling user-controllable dereverberation~\cite{raoLowComplexityNeuralSpeech2025}.

A fourth class of methods consists in generative approaches.
While previously-enumerated deep learning models are discriminative and require pairs of dry and wet data, generative approaches use a prior that is pre-trained on dry data only.
At inference, they model reverberation using the RIR itself with a diffusion-based prior~\cite{lemercierDiffusionPosteriorSampling2023}, or leverage autoregressive models of reverberation combined with
Recurrent VAE~\cite{wangRVAEEMGenerativeSpeech2024}, 
or diffusion~\cite{lemercierUnsupervisedBlindJoint2025}
priors.
Note that some methods make no assumption on the reverberation model and %just 
only
use the noisy signal~\cite{richterSpeechEnhancementDereverberation2023} or a pre-trained dereverberation model~\cite{murataGibbsDDRMPartiallyCollapsed2023} to start the reverse diffusion process.

All the aforementioned techniques require dry signals at training.
There are only a few approaches that are based on models 
trained uniquely using reverberant signals.
For instance, MetricGAN-U~\cite{fuMetricGANUUnsupervisedSpeech2022} avoids the use of any explicit reverberation model, instead relying on supervision from a pre-trained metric to guide the dereverberation process. 
Herein, we refer to this method as metrics-based dereverberation.
In addition, AR models of reverberation can be trained in an unsupervised fashion, including a neural version of WPE ~\cite{petkovUnsupervisedLearningApproach2019}, and USDNet~\cite{wangUSDnetUnsupervisedSpeech2024} 
{but only demonstrating marginal improvement, if any, compared to WPE.}

In our previous work~\cite{bahrmanHybridModelWeaklySupervised2025}, we demonstrated that a dereverberation model could be trained with weak supervision based on reverberation parameters. This reverberation-based weak supervision outperformed the metrics-based supervision used in MetricGAN-U. 
Building on this, the current work seeks to bridge the gap between the autoregressive model parameterized by weak supervision, as presented in~\cite{bahrmanHybridModelWeaklySupervised2025} (which offers improved performance at the cost of reliance on a dataset of reverberation parameters), and the unconstrained AR model (which exhibits suboptimal performance).
We study how our proposed hybrid unsupervised training method interacts with several supervised data-driven dereverberation models, and compare it to unconstrained autoregressive approaches.

\section{Reverberation model}~\label{sec:notations}

Assuming fixed source and microphone positions, the monaural reverberant observation $y$ can be modeled as the convolution of the anechoic source signal $s$ with the room impulse response (RIR) $h$, contaminated by additive noise $\epsilon$. The resulting signal is expressed as:
\begin{equation}
y(n) = (s \star h)(n) + \epsilon(n), \label{eq:noisy_formulation_time_domain}
\end{equation}
where $\star$ denotes the linear convolution operator and $n$ is the discrete time index.

\subsection{Room Impulse Response Model}

The room impulse response $h$ is typically decomposed into two components: 
the direct-path response $h_d$, corresponding to the initial peak of the RIR, and the subsequent reverberant tail $h_r$, which begins after a delay of $n_d$ samples. 
This delay
accounts for the temporal extent of the direct path. 
In practice, the initial delay of the RIR is ignored as it only causes a temporal misalignment which is unperceptible in subjective evaluation but causes a notable degradation of most objective metrics.
Therefore, $n_d$ is fixed at \qty{2.5}{\milli\second}, corresponding
to 40 samples at a sampling rate of \qty{16}{\kHz}~\cite{eatonEstimationRoomAcoustic2016}.

A key descriptor of room acoustics is the direct-to-reverberant ratio (DRR), which quantifies the energy balance between the direct path and the reverberant tail. As defined 
% in~\cite[Sec.~2.4.2]{naylorSpeechDereverberation2010}, 
in~\cite{naylorModelsMeasurementEvaluation2010},
the DRR is given by:
\begin{equation}
\mathrm{DRR} = 10\log_{10}\left( \frac{\sum_{n=0}^{n_d} h^2(n)}{\sum_{n=n_d+1}^{\infty} h^2(n)} \right) \si{\dB},
\end{equation}
The DRR is widely used as a quantitative measure for evaluating and modeling reverberant conditions.

Another fundamental parameter is the reverberation time, denoted $\rtsixty$, which corresponds to the time required for the sound energy to decay by \qty{60}{\dB}. Under idealized conditions~\cite{naylorModelsMeasurementEvaluation2010}, $\rtsixty$ can be estimated from the slope of the energy decay curve (EDC), introduced by Schroeder~\cite{schroederComplementaritySoundBuildup1966}. 

The DRR and the $\rtsixty$ are sufficient parameters to characterize Polack's statistical model of reverberation~\cite{polackTransmissionLenergieSonore1988}. This model assumes that the reverberant tail of an RIR can be modeled as an exponentially decaying stochastic process. Specifically, the reverberant component $h_r$ is defined as:
\begin{equation}
h_r(n) = b(n) e^{-n/\tau}, \label{eq:polack}
\end{equation}
where $b(n) \sim \mathcal{N}(0, \sigma^2)$ denotes a zero-mean white Gaussian noise process, and the decay constant $\tau$ is related to the $\rtsixty$ and the sampling frequency $f_s$ by:
\begin{equation}
\tau = \frac{\rtsixty f_s}{3 \ln(10)}. \label{eq:tau}
\end{equation}

This statistical model can be combined with 
a model of the direct-path as a delayed impulse
in order to simulate a full RIR 
using only a small set of physically meaningful acoustic parameters.

\subsection{Convolution in the Time-Frequency Domain}\label{sec:convolution_in_tf_domain}

In the noiseless case, the linear time-invariant filtering in time-domain in Eq.~\eqref{eq:noisy_formulation_time_domain} can be equivalently expressed in the %GR: the short-time Fourier transform (
$\STFT$ domain
 using an inter-frame and inter-band convolution operator denoted $\mathcal{C}$, defined in~\cite{avargelSystemIdentificationShortTime2007} as:
% , under the assumption of time-invariant linear filtering. As shown in,
\begin{align}
\bm{Y}_{f,t}
=
\mathcal{C}(\bm{S},h)_{f,t}
\triangleq
\sum_{f^{\prime}=0}^{F-1}\sum_{t^{\prime}=0}^{\min(t; T_h)} \mathcal{H}_{f,f^{\prime},t^{\prime}} S_{f^{\prime},t-t^{\prime}}
, 
\label{eq:full_convolution}
\end{align}
where $\bm{Y} \triangleq \{Y_{f,t}\}_{f,t=0}^{F-1, T_y-1} \in \mathbb{C}^{F \times T_y}$ denotes the $\STFT$ coefficients of the reverberant signal at frequency $f\thineq 0,\dots,F-1$ and time $t\thineq 0,\dots,T_y-1$, 
$\bm{S} \triangleq \{S_{f,t}\}_{f,t=0}^{F-1, T_s-1}\in \C^{F\times T_s}$ is the corresponding $\STFT$ of the dry signal,
and
$\mathcal{H} \triangleq \{\mathcal{H}_{f, f^\prime, t}\}_{f,f^\prime, t=0}^{F-1,F-1,T_h-1} \in \C^{F \times F \times T_h}$
represents the time-frequency convolution kernel induced by the RIR, capturing both spectral and temporal spread.
The convolution kernel $\mathcal{H} $ is derived from the time-domain RIR $h \in \mathbb{R}^{N_h}$ by~\cite{avargelSystemIdentificationShortTime2007}:
\begin{equation}
    \mathcal{H}_{f, f', t'}
    =
    \sum_{m=-N+1}^{N-1} h(t'L - m) W_{f,f^\prime}(m), \label{eq:big_H}
\end{equation}
where $N$ denotes the $\STFT$ window length, $L$ the hop size, and
\begin{equation}
     W_{f,f^\prime}(m)=\frac{1}{F}\sum_{n=0}^{N-1}
    g_s(n + m) g_a(n) e^{\frac{j2\pi (f^\prime (n+m) - fn)}{F}} \label{eq:big_window} 
\end{equation}
Here, $g_s$ and $g_a$ denote the synthesis and analysis window functions, respectively.
The formulation of Eq.~\eqref{eq:full_convolution} enables the integration of time-domain models of reverberation into time-frequency processing frameworks, which is particularly advantageous for both estimation and learning-based approaches.

\section{The U-DREAM model}\label{sec:method}

In this section, we derive the formulation  of our proposed Unsupervised Dereverberation system guided by a REverberAtion Model (U-DREAM).

\subsection{Problem formulation}\label{sec:problem_formulation}

The noisy time-domain formulation of Eq.~\eqref{eq:noisy_formulation_time_domain} can be expressed in the time-frequency domain using Eq.~\eqref{eq:full_convolution} by introducing the $\STFT$ of the noise term:
\begin{equation}
    \bm{Y}_{f,t} = \mathcal{C}(\bm{S},h)_{f,t} + \mathcal{E}_{f,t},
    \label{eq:convolutive_model_stft_with_noise}
\end{equation}
where 
$\bm{\mathcal{E}} \triangleq \{\mathcal{E}_{f,t}\}_{f,t=0}^{F-1, T_s-1}\in \C^{F\times T_y}$ is the $\STFT$ of the additive noise.
In this work, we assume that all $\mathcal{E}_{f,t} \sim \N_\C(0,\nu^2)$ are independent and identically distributed (i.i.d.) complex Gaussian variables, and that the dry signal $\STFT$ is deterministic.
Under such assumption, 
\begin{equation}
    \bm{Y}_{f,t} \mid h; \bm{S},\Theta \sim \N_\C(\mathcal{C}(\bm{S},h)_{f,t}, \nu^2)
    \label{eq:conditional_proba_of_y}
\end{equation}

Introducing the finite RIR $h$ as a random
vector conditioned only on the acoustic parameters $\Theta$, the total probability distribution of $\bm{Y}$ is:
\begin{equation}
    p(\bm{Y}; \bm{S}, \Theta) = \int p(\bm{Y} \mid h; \bm{S},\Theta) p(h ; \Theta) dh.
\end{equation}
The integral can be expressed as an expectation with respect to the distribution $ p(h;\Theta)$,
and its negative log-likelihood is:
\begin{align}
     -\log p(\bm{Y}; \bm{S}, \Theta) = - \log \E_{p(h\mid\Theta)} \left[ p(\bm{Y} \mid h;\bm{S},\Theta) \right].
\end{align}
Jensen's inequality can be applied to obtain an upper bound of the generally intractable negative log-likelihood of the expectation:
\begin{align}
    - \log p(\bm{Y};\bm{S},\Theta) &\leq \E_{p(h \mid \Theta)} \left[ -\log  p(\bm{Y}\mid h;\bm{S},\Theta) \right].
    \label{eq:jensen_relaxation}
\end{align} 
Finally, replacing the conditional probability distribution of $\bm{Y} \mid h; \bm{S},\Theta$ from Eq.~\eqref{eq:conditional_proba_of_y} yields:
\begin{align}
- \log p(\bm{Y};\bm{S},\Theta) &\leq
     % & \E_{p(h\mid\Theta)} \left[ -\log  p(\bm{Y}\mid\bm{S},h) \right] \\
    & \E_{p(h\mid\Theta)} \left[ \norm{\bm{Y} - \mathcal{C}({\bm{S}}, h)}^2_F \right] + C,
    \label{eq:loss_relaxed_with_epsilon}
\end{align}
where $\norm{\cdot}_F $ denotes the Frobenius norm, and $C=-F T_y \log(\pi \nu^2)$ is a constant with respect to $\bm{S}$ and $\Theta$.

The task of dereverberating a signal using an acoustic model can be formulated as a maximum likelihood estimation problem, where the goal is to jointly estimate both the $\STFT$ of the dry speech signal $\bm{S}$ and the acoustic parameters $\Theta$ of a parametric reverberation model, given the observed reverberant $\STFT$ $\bm{Y}$. Formally, this is expressed as:
\begin{equation}
    \argmax_{\bm{S},\Theta} p(\bm{Y}; \bm{S},\Theta).
\end{equation}
In this work, we solve a relaxed version of the problem, where we minimize the upper bound of the negative log-likelihood obtained from Eq.~\eqref{eq:loss_relaxed_with_epsilon}:
\begin{equation}
    \hat{\bm{S}}, \hat{\Theta} = \argmin_{\bm{S},\Theta} \E_{p(h\mid\Theta)} \left[ \norm{\bm{Y} - \mathcal{C}({\bm{S}}, h)}_F^2 \right].
    \label{eq:relaxed_argmin}
\end{equation}

\subsection{Overview of the method}

To address the ill-posed nature of solving Eq.~\eqref{eq:relaxed_argmin} in the monaural case, we introduce a model-based deep learning framework.
% , detailed in the next subsections.
Specifically, we propose to replace the direct optimization of $\bm{S}$ and $\Theta$ with two trainable, model-based mappings: a dereverberation module $\mathcal{D}_{w_D}: \bm{Y} \mapsto \hat{\bm{S}}$ and an acoustic analyzer $\mathcal{A}_{w_A}: \bm{Y} \mapsto \hat{\Theta}$,
where $w_D$ and $w_A$ denote the weights that parametrize each model (sometimes omitted for sake of clarity).

\begin{figure}
    \centering
    \includegraphics[width=.7\linewidth]{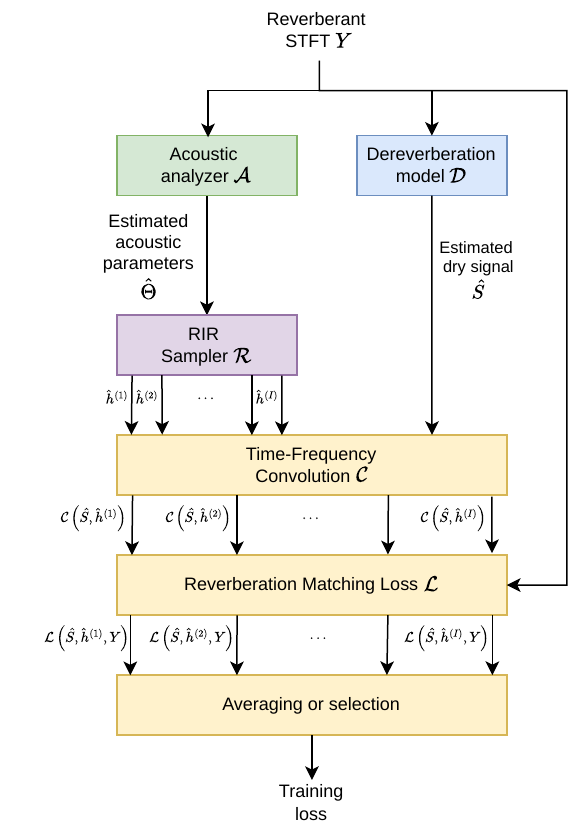}
    \caption{Overview of the proposed method}
    \label{fig:overview}
\end{figure}

Substituting the optimization of $\bm{S}$ and $\Theta$ with the optimization of $w_D$ and $w_A$, the problem in Eq.~\eqref{eq:relaxed_argmin} becomes:
\begin{equation}
    w_D, w_A \in \argmin_{w_D,w_A} \E_{p \left( h\mid\mathcal{A}_{w_A}(\bm{Y}) \right)} \left[ \norm{\bm{Y} - \mathcal{C} \left(\mathcal{D}_{w_D}(\bm{Y}), h \right)}_F^2 \right].
    \label{eq:deep_lerning_joint_weights_argmin}
\end{equation}
We propose to solve this equation by training both models $\mathcal{A}$ and $\mathcal{D}$ 
using Stochastic Gradient Descent (SGD) on a dataset of reverberant signals. 
The overall forward of the proposed framework is illustrated in Fig.~\ref{fig:overview} and summarized as follows:
given a reverberant signal $\bm{Y}$, the dereverberation module $\mathcal{D}$ produces an estimated signal $\hat{\bm{S}}$. 
Simultaneously, the acoustic analyzer $\mathcal{A}$ (see in Section~\ref{sec:acoustic_analyzer}) estimates the corresponding acoustic parameters $\hat{\Theta}$ from $\bm{Y}$.
To approximate the expectation $\E_{p(h\mid \hat{\Theta})}$ by Monte-Carlo sampling, one or more RIRs $\hat{h} \in \mathbb{R}^{N_h}$ are drawn from a reverberation sampler $ \mathcal{R} $ 
(See Section~\ref{sec:rir_sampler}).
Each RIR  $\hat{h}$ is convolved with the estimated dry signal $\hat{\bm{S}}$ via the operator 
$\mathcal{C}$, producing one or more estimated reverberant $\STFT$s $\hat{\bm{Y}}$. 
The optimization objective is expressed through a reverberation matching loss function $\mathcal{L}$ (described in Section~\ref{sec:loss_variants_and_gradnorm}), which quantifies the distance between the estimated reverberant $\STFT$ $\hat{\bm{Y}}$ and the observed reference $\bm{Y}$.

A critical aspect of this framework is the potential trivial solutions if both $\mathcal{A}_{w_A}$ and $\mathcal{D}_{w_D}$ are trained jointly from scratch. 
Specifically, the acoustic analyzer $\mathcal{A}$ could converge to predicting acoustic parameters corresponding to an anechoic environment (e.g., low $\rtsixty$ or high $\DRR$), enabling the dereverberation module $\mathcal{D}$ to simply learn an identity mapping, thereby bypassing the intended dereverberation process. 
To mitigate this issue, we adopt a two-stage training strategy. The acoustic analyzer  $\mathcal{A}$ is first pre-trained using available supervised data. Once  $\mathcal{A}$ has been pre-trained, we proceed to train $\mathcal{D}$ while keeping $\mathcal{A}$ frozen.
This staged approach is motivated by the relative difficulty of the two tasks: predicting $\Theta$ is inherently easier than estimating $\bm{S}$, and 
%the acoustic analyzer can typically be trained with significantly less data than the dereverberation  module.
the acoustic analyzer can typically yield a satisfactory reverberation matching loss with significantly less data than the dereverberation module. 

At inference for the dereverberation task, only the dereverberation model is used.
Hence, the number of parameters, as well as the
computational complexity and memory footprint are the same
as if the model had been trained with full supervision.

Each module is detailed in the next sections.

\subsection{Reverberation sampler \mf{$ \mathcal{R} $}}\label{sec:rir_sampler}

The RIR sampler is responsible for generating one or more RIRs $h$ sampled from the conditional distribution $p(h\mid\Theta)$, where $\Theta$ represents the acoustic parameters.
The goal is to synthesize an RIR whose key characteristics, such as $\rtsixty$ and $\DRR$, match those  of the original setting in which the reverberant signal was recorded.
In this work, the RIR sampler is based on Polack's model, detailed in Eq.~\eqref{eq:polack}.
For synthetic data, a slight modification to this model is introduced to better account for characteristics observed in simulated RIRs.
In particular, Image source model (ISM)-based RIR generators such as Pyroomacoustics~\cite{scheiblerPyroomacousticsPythonPackage2018} often produce RIRs that exhibit nonzero energy at the zero frequency. 
To better match such RIRs, we modify the noise component of Polack's model. Specifically, when encountering non-centered RIRs, we match the zero-frequency energy of the ground-truth RIR while maintaining a flat spectral response in the other bands by drawing the noise term $b(n)$ from a half-normal distribution $\abs{\mathcal{N}(0, \sigma^2)}$.
Preliminary experiments demonstrated that this modification reduces the loss defined in Eq.~\eqref{eq:loss_relaxed_with_epsilon} under oracle conditions (i.e., oracle dry speech $\bm{S}$ and acoustic parameters $\Theta$, and no additive noise).
To further simplify the model and ensure proper scaling, all direct-path energy is concentrated at the peak of the RIR, with normalization applied such that the peak value is $1$.
The resulting reverberation sampler $ \mathcal{R} $ is defined as:
\begin{equation}
     \mathcal{R} (\Theta)(n) = \begin{cases}
        b(n) e^{-\frac{3 \ln(10)}{\rtsixty f_s} n } & \text{if } n > n_D\\
        1 & \text{if } n = 0\\
        0 & \text{otherwise},
    \end{cases}
    \label{eq:reverb_sampler}
\end{equation}
where $b(n) \sim \mathcal{N}(0, \sigma^2)$ for measured and $b(n) \sim \abs{\mathcal{N}(0, \sigma^2)}$ 
for synthetic ground-truth RIRs respectively.

According to Polack's model, the reverberant energy $E_R$ is given by:
% \begin{align}
%     E_R &= \int_{n_D}^{+\infty} \sigma^2 e^{-2t/\tau} dt\\
%     &= \sigma^2 \frac{\tau}{2} e^{-2 n_D / \tau},
%     \label{eq:reverberant_polack_energy}
% \end{align}
\begin{equation}
    E_R = \int_{n_D}^{+\infty} \sigma^2 e^{-2t/\tau} dt
    = \sigma^2 \frac{\tau}{2} e^{-2 n_D / \tau},
    \label{eq:reverberant_polack_energy}
\end{equation}
where $\tau$ is defined in Eq.~\eqref{eq:tau}

Assuming the direct-path energy is normalized to $1$, $\sigma$ can be computed from the target $\DRR$ as:
\begin{equation}
    \sigma=\sqrt{\frac{2 e^{2 n_D/\tau}}{\tau \DRR}}
\end{equation}

In addition to this probabilistic sampler, we introduce an "oracle" reverberation synthesis module used for comparison in the experimental section. In this oracle setting, the distribution $p(h\mid\Theta)$ is replaced by a Dirac measure $\delta_{\Theta}(h)$, where $h$ corresponds exactly to the true RIR used to generate the reverberant signal in the training set.

The acoustic parameters $\Theta$ required in the reverberation synthesis module are estimated using the acoustic analyzer described hereunder.

\subsection{Acoustic Analyzer $\mathcal{A}$}\label{sec:acoustic_analyzer}

The acoustic analysis module estimates the reverberation parameters $\Theta$, namely $\rtsixty$ and $\DRR$, which are used to guide the reverberation synthesis process described in Section~\ref{sec:rir_sampler}.
We distinguish between two modes of analysis:
The \textit{Non-blind} case in which the ground-truth RIR $h$ is known, and the \textit{blind} case, 
where only the reverberant signal $\bm{Y}$ is available.

\subsubsection{Non-blind analysis}

in the non-blind setting, acoustic parameters are derived directly from the RIR.
We follow the procedure described in~\cite{naylorModelsMeasurementEvaluation2010}, based on linear regression of the energy decay curve $\EDC$.
However, measured RIRs, often exhibit noise in the late reverberation tail, which can bias estimates of $\rtsixty$ and $\DRR$. 
To mitigate this issue, we restrict the analysis to the dynamic range in which Polack's model is considered to be valid, namely between  $\qty{-5}{\decibel}$ and $\qty{-25}{\decibel}$, as
outside this range, the measured $\EDC$ and the theoretical model diverge.
The energy on this range, noted $E_{25}^{5}$,  is expected to follow:
\begin{equation}
    E_{25}^{5}= \sigma^2 \frac{\tau}{2} \left( e^{-2 T_{5}/\tau} - e^{-2 T_{25} /\tau} \right),
    \label{eq:energy_minus_5_minus_25_from_sigma}
\end{equation}
where $T_{5}$ and $T_{25}$ represent the time it takes for the EDC to decrease by $\qty{5}{\decibel}$ and $\qty{25}{\decibel}$ respectively 
The $\rtsixty$ is first computed from the slope of the EDC, which is independent of the total energy of the RIR. Then, using the measured value of $E_{25}^{5}$
and Eq.~\eqref{eq:energy_minus_5_minus_25_from_sigma}, the parameter $\sigma$ is derived, leveraging the relation between $\tau$ and $\rtsixty$ given in Eq.~\eqref{eq:tau}.

Finally, $\sigma$ can be directly reused in the reverberation sampler $\mathcal{R}$ detailed in Eq.~\eqref{eq:reverb_sampler}, or to compute the $\DRR$ using Eq.~\eqref{eq:reverberant_polack_energy}.
Note that this approach excludes the contribution of late reverberation noise to the estimated $\DRR$, thereby improving alignment with the synthesis model.

\subsubsection{Blind \mf{Analysis}}

In the blind case, both $\rtsixty$ and $\DRR$ must be estimated from the reverberant $\STFT$ $\bm{Y}$.
For the $\rtsixty$ estimation, we adopt the method proposed in~\cite{prego_2015_blind}. 
This approach computes local energy decay in specific regions of $\bm{Y}$. A polynomial mapping is then used to match the median $\rtsixty$ found in the time-frequency domain to its corresponding time-domain value. 
Based on preliminary experiments measuring the accuracy of $\rtsixty$ estimation, we employ a second-order polynomial, instead of the first-order polynomial used in the original study.
The three polynomial coefficients are tuned on a very small calibration dataset
of \num{100} pairs $(\bm{Y}, \rtsixty)$.

For the $\DRR$ estimation, we adopt the BiLSTM-based model of~\cite{mackSingleChannelBlindDirecttoReverberation2020}, which predicts two time-frequency domain energy masks.
The ratio between the masked spectrograms is used to estimate the $\DRR$.

\subsection{Reverberation Matching Loss and variants}\label{sec:loss_variants_and_gradnorm}

This section details the \textit{reverberation matching} loss, which quantifies the distance between the ground-truth reverberant $\STFT$ $\bm{Y}$ and the estimated reverberant $\STFT$ produced by convolving the estimated dry signal $\hat{\bm{S}}$ with an RIR sampled by the reverberation sampler $\hat{h}$.

The loss inside the expectation in Eq.~\eqref{eq:relaxed_argmin} assumes i.i.d. complex additive noise on each bin of the reverberant $\STFT$ $\bm{Y}$. We denote this loss as:
\begin{equation}
    \L_\C (\bm{Y},\hat{\bm{S}},\hat{h}) \triangleq \norm{\bm{Y} - \mathcal{C}({\hat{\bm{S}}}, \hat{h})}_F^2.
    \label{eq:loss_complex}
\end{equation}
We also consider an additional term, denoted $\L_{\text{MAG}}$, that computes a distance in the log-magnitude space as in~\cite{schwarMultiScaleSpectralLoss2023}:
\begin{equation}
    \L_{\text{MAG}}(\bm{Y},\hat{\bm{S}},\hat{h})
    \triangleq
    \norm{\log \frac{ 1+ \abs{\bm{Y}}}{1+ \abs{\mathcal{C}(\hat{{\bm{S}}}, \hat{h})} }}_F^2.
    \label{eq:loss_magnitudes}
\end{equation}
The total per-sample reverberation matching loss $\L$ sums these two terms 
using a weight $\alpha > 0$
% as the weighted sum % with $\alpha > 0$
\begin{equation}
    \L = \L_\C + \alpha  \L_{\text{MAG}}.
    \label{eq:sum_losses}
\end{equation}
Preliminary experiments showed that adding the log-magnitudes loss term $\L_{\text{MAG}}$ enabled faster convergence of our proposed framework.
% Combining these two losses with the expectation in Eq.~\eqref{eq:deep_lerning_joint_weights_argmin}, our training loss is the expectation
Using Eq. \eqref{eq:deep_lerning_joint_weights_argmin}, \eqref{eq:loss_complex}-\eqref{eq:sum_losses}, and considering the linearity of the expectation, it can be shown that our training loss is:
\begin{equation}
\E_{p(\hat{h} \mid \mathcal{A}_{w_A}(\bm{Y}))} \L(\bm{Y},\hat{\bm{S}},\hat{h}) .
    \label{eq:training_loss_with_complex_and_log}
\end{equation}
We now detail some sampling strategies used to compute this expectation.

\subsubsection{Loss variants}\label{sec:loss_variants}

The expectation in Eq.~\eqref{eq:training_loss_with_complex_and_log} is computed via Monte Carlo sampling. Depending on the sampling strategy adopted, the loss has different physical interpretations regarding the underlying reverberant scene.
In particular, multiple draws of the RIR from the sampler $ \mathcal{R} (\Theta)$ can be viewed as simulating different virtual microphone positions within a room characterized by $\Theta$.
Variability between these draws reflects potential variations in the observed reverberant signal due to changes in microphone placement, even though the target signal $\bm{Y}$ corresponds to a single physical microphone. In this work, we consider three sampling strategies:
\begin{itemize}
    \item \textbf{Single}: 
    A single RIR $\hat{h}$ is drawn from $ \mathcal{R} (\Theta)$ for each reverberant spectrogram $\bm{Y}$. Across different epochs, different RIR draws may be associated with the same $\bm{Y}$. Physically, this corresponds to simulating one virtual microphone in a room with parameters $\Theta$, while acknowledging that the exact microphone position may not match that of $\bm{Y}$.
    The loss is computed as:
    \begin{equation}
        \L_{\text{single}}
    =  \L(\bm{Y},\hat{\bm{S}},\hat{h}).
    \end{equation}
    
    \item \textbf{Average}: 
    Multiple RIRs $\{ \hat{h}^{(i)} \}_{i=1}^I$ are drawn from $ \mathcal{R} (\Theta)$, and the loss is computed as the mean across these draws. This corresponds to simulating $I$ virtual microphones in the same room and averaging their contribution to the loss. In our experiments, $I = 10$ draws is used:
    \begin{equation}
        \L_{\text{avg}}
    =  \frac{1}{I} \sum_{i=1}^{I}
    \L(\bm{Y},\hat{\bm{S}},\hat{h}^{(i)}).
    \end{equation}

    \item \textbf{Best}:
    As in the "Average" strategy, multiple ($I=10$) RIRs are drawn, but only the draw yielding the lowest loss is used for backpropagation. This can be interpreted as searching for the virtual microphone position in the room that best explains $\bm{Y}$, and encouraging the model to match this optimal configuration:
       \begin{equation}
        \L_{\text{best}}
    =  \min_i
    \L(\bm{Y},\hat{\bm{S}},\hat{h}^{(i)}).
    \end{equation}
    
\end{itemize}

\subsubsection{Balancing the Loss Terms}

To ensure that $\L_\C $ and $ \L_{\text{MAG}}$ contribute equally during training, we adopt the GradNorm method~\cite{chenGradNormGradientNormalization2018} to automatically adjust the weight $\alpha$
\lreplace{. GradNorm balances the norm of the gradients
\mfcmt{la norme de Frobenius ou autre ? Juste pour être précis}
}{
such that the Frobenius norm of the gradients of both losses with respect to the model weights are equal.
}
In the case of the $\L_\text{Best}$ strategy, $\alpha$ and the selected index of the optimal RIR draw $i$ are interdependent, which results in a bi-level optimization problem. To circumvent this issue, we approximate the GradNorm method by solving for each sampled RIR $i$:
\begin{equation}
    \norm{\frac{\partial \L_\C(\bm{Y},\hat{\bm{S}},\hat{h}^{(i)})}{\partial \mathcal{C}(\hat{\bm{S}}, \hat{h})}} = \norm{ \alpha_i \frac{\partial \L_\text{MAG}(\bm{Y},\hat{\bm{S}},\hat{h}^{(i)})}{\partial \mathcal{C}(\hat{\bm{S}}, \hat{h})}}
    .
\end{equation}
We then select the RIR draw $\hat{i}$ that yields the lowest loss, and backpropagate gradients only through $\mathcal{C}(\hat{\bm{S}},\hat{h}^{(i)})$ with the corresponding optimal weight $\alpha_{\hat{i}}$

\subsection{Training-less variant}\label{sec:trainingless_variant}

An alternative formulation of Eq.~\eqref{eq:relaxed_argmin} is also considered, wherein only the acoustic analyzer $\mathcal{A}_{w_A}$ is employed, without the need for a dereverberation module.
This training-less variant assumes that $\mathcal{A}_{w_A}$ has been pre-trained. Given a reverberant signal $\bm{Y}$ and fixed analyzer parameters $w_A$, the following optimization problem is solved directly for $\hat{\bm{S}}$:
\begin{equation}
    \hat{\bm{S}} = \argmin_{\bm{S}} \E_{p(h\mid\mathcal{A}_{w_A}(\bm{Y}))} \left[ \norm{\bm{Y} - \mathcal{C}({\bm{S}}, h)}_F^2 \right] \label{eq:argmin_trainingless}
\end{equation}
Note that a closed-form solution of the training-less problem exists in a very simple case: The time-domain formulation of Eq.~\eqref{eq:argmin_trainingless}, with RIRs synthesized using Polack's model applied on Gaussian Noise admits closed-form solution that can be computed from acoustic parameters\footnote{See \url{louis-bahrman.github.io/UDREAM/proofs.pdf} for details.}.
Nevertheless, the training-less variant enables our proposed framework to be used with any reverberation model, in case a closed form of the solution might not exist.
This is for instance the case in our adaptation of Polack's model to simulate synthetic RIRs.
This variant is named \textit{training-less} because it does not require a dereverberation module $\mathcal{D}_{w_D}$ to be fitted on a large dataset but rather directly optimizes the dry $\STFT$ $\bm{S}$ on a per-sample basis given the output of the acoustic analyzer.

\section{Experimental setup}\label{sec:experimental_setup}

We tested our proposed dereverberation guided by a reverberation model framework under several supervision paradigms on datasets of both real and synthetic RIRs. This section presents our experimental protocol.

\subsection{Experiments summary}

We conduct four experiments to assess the performance of our framework on two tasks: acoustic parameter estimation and dereverberation. We explore several supervision scenarios, including strong supervision, weak supervision and unsupervised dereverberation.

\subsubsection{Strong supervision for dereverberation}

In this experiment, the dereverberation module $\mathcal{D}_{w_D}$ is trained using full supervision, where the acoustic parameter set $\Theta$ corresponds to the ground-truth RIR $h$ and $p(h\mid\Theta)$ is modeled as a Dirac distribution centered on $h$ (as described in Section~\ref{sec:rir_sampler}).
Under this setting, the expectation in Eq.~\eqref{eq:deep_lerning_joint_weights_argmin} simplifies, yielding the following optimization objective:
\begin{equation}
    \argmin_{w_D} \L(\bm{Y},\mathcal{D}_{w_D} (\bm{Y}),{h})\label{eq:strong_rir_supervision},
\end{equation}

We compare this supervision to the strong supervision of the ground truth dry signal $\bm{S}$, with the original training loss used for each dereverberation model.

\subsubsection{Weak supervision for dereverberation}
In this setting, we provide oracle acoustic parameters $\Theta$ to guide training, while sampling RIRs $\hat{h} \sim p(h \mid \Theta)$.
 The optimization objective is given by:
\begin{equation}
    \argmin_{w_D} \E_{p(\hat{h}\mid\Theta)} \L(\bm{Y},\mathcal{D}_{w_D} (\bm{Y}),\hat{h})
\end{equation}
We compare the three different Monte Carlo estimation strategies described in Section~\ref{sec:loss_variants} to assess their impact in this weakly-supervised training regime.

\subsubsection{Acoustic parameter estimation with various supervision}

In this experiment, we evaluate the acoustic analyzer $\mathcal{A}_{w_A}$ trained under different supervision settings. The task is to estimate acoustic parameters $\Theta$ from reverberant observations $\bm{Y}$, with access to either the corresponding clean speech $\bm{S}$ or acoustic parameters $\Theta$. 
We focus on the task of $\DRR$ estimation ($\Theta = \{\DRR\}$).
In the \textit{Reverberation Matching} (RM) supervision setting, we leverage pairs $(\bm{Y}, \bm{S})$, optimizing the following objective:
\begin{equation}
    \argmin_{w_A} \E_{p(\hat{h}\mid\mathcal{A}_{w_A}(\bm{Y}))} \left[ \L(\bm{Y},{\bm{S}},\hat{h}) \right] 
    \label{eq:acoustic_parameter_estimation_reverb_matching_loss}
\end{equation}
In the \textit{Parameter matching} (PM) setting, only pairs $(\bm{Y}, \Theta)$ are provided, and training proceeds by minimizing:
\begin{equation}
    \argmin_{w_A} \norm{\mathcal{A}_{w_A}(\bm{Y}) - \Theta}_2^2
    \label{eq:acoustic_parameter_estimation_parameter_matching_loss}
\end{equation}
We also study the relationship between model performance and the amount of available training data. Based on results of the previous experiment, we use the \textit{single} loss variant from Section~\ref{sec:loss_variants} for this experiment, as loss variant choice was found to have negligible impact here.

\subsubsection{Unsupervised dereverberation guided by a reverberation model}
Finally, we explore an unsupervised dereverberation scenario. In this case, we reuse the pre-trained acoustic analyzer $\mathcal{A}_{w_A}$ obtained from the previous experiment, and perform dereverberation by directly optimizing $\bm{S}$ as:
\begin{equation}
    \argmin_{w_D} \E_{p(\hat{h}\mid\hat{\Theta})} \L(\bm{Y},\mathcal{D}_{w_D}(\bm{Y}),\hat{h})
\end{equation}
where $\hat{\Theta}$ is determined using the pre-trained acoustic analyzer $\mathcal{A}_{w_A}$.
This regime is considered unsupervised, as no ground-truth clean speech $\bm{S}$, RIR $h$, or acoustic parameters $\Theta$ are used during dereverberation.

\subsection{Datasets}

We evaluate our procedure on both synthetic and real reverberation. We consider 2 datasets: EARS-Reverb and EARS-Synth.

\subsubsection{EARS-Reverb}

The EARS (Expressive Anechoic Recordings of Speech)~\cite{richterEARSAnechoicFullband2024} dataset is composed of high-quality dry speech signals recorded from various speakers and diverse content.
We use its dereverberation benchmark,
EARS-Reverb, generated by convolving anechoic speech from
EARS with RIRs sampled from publicly-available datasets.
In this dataset, 
the beginning of each RIR is cut off up to the index with the highest amplitude,
to avoid a time delay between the reverberant and clean speech signal.
This dataset is provided at a sampling rate of \qty{48}{\kHz}. 
Our task being dereverberation at \qty{16}{\kHz}, 
we resample clean, reverberant signals, and RIRs to this sampling rate.

\subsubsection{EARS-ISM}

We consider a simpler reverberant dataset, composed of anechoic audio from EARS and simulated reverberation synthesized using the Image-Source Method in Pyroomacoustics~\cite{scheiblerPyroomacousticsPythonPackage2018}.
The simulated RIR dataset consists of
32,000 RIRs drawn from 2000 rooms simulated using the image source method implemented in the pyroomacoustics library \cite{scheiblerPyroomacousticsPythonPackage2018}.
Room dimensions and RT60 are uniformly
sampled in the respective ranges of $[5,10] \times [5,10] \times [2.5,4]~\text{m}^3$, and $[0.2,1.0]$~s.
The source-microphone distance is uniformly distributed in $[0.75, 2.5]$~m, 
and both source and microphone are at least $50$~cm from the
walls.
At training time, we use a dynamic mixing procedure
consisting of randomly selecting a dry signal and RIR pair. 
In order to align the dry signal target and the direct-path, the samples before the direct path are discarded and it is normalised (so that the direct-path is of amplitude 1). This does not change
the RIR distribution and compensates for the delay induced by
the direct path to match the RIR synthesis procedure.

\subsubsection{Out-Of-Domain}
We also evaluate the generalization performance on models trained on synthetic RIRs and tested on real RIRs.

\subsection{Speech models}

Our proposed dereverberation supervision paradigms are, in theory, adaptable to any speech model.
In practice, such deep-learning-based speech models are very diverse and their performance is variable.
Indeed, it is relevant to consider how the reverberation-aware training interacts with various dry speech priors.
We present dereverberation results for three deep-learning-based speech models:
\begin{itemize}
    \item BiLSTM~\cite{weningerSpeechEnhancementLSTM2015}: 
    This model consists of a 2-layer
bidirectional LSTM model followed by a linear layer,
processing sub-bands of the degraded STFT magnitudes in a recursive manner.
    Since this model is only able to process magnitude masks, it can be considered as phase-agnostic, and the underlying speech model is that both dry and reverberant speech are Gaussian complex noise, with circular invariance.
    \item FSN~\cite{haoFullsubnetFullBandSubBand2021}: This more powerful LSTM \mf{called FullSubNet} is capable of recursive cross-band processing and, since it is able to generate a complex-valued mask, it can be considered as a phase-aware model. This model has also been used in reverberation-aware training~\cite{zhouSpeechDereverberationReverberation2023}.
    \item TFL~\cite{saijoTFLocoformerTransformerLocal2024}: TF-Locoformer represents a state-of-the-art model for speech enhancement \mf{and dereverberation}, which is very powerful and expressive thanks to its self-attention module capable of global modeling and convolution handling local modeling.
\end{itemize}
These models are a representative set of DNN-based dereverberation methods.

\subsection{Misc settings}

During training, 4-second excerpts are processed at $\qty{16}{\kHz}$. 
$\STFT$ processing is done using a 512-sample Hann window with an overlap of $\qty{50}{\percent}$. 
All DNNs are trained using Adam optimizer, with a learning rate of $\num{e-4}$, and early stopping based on the STOI metric on a validation set is used.
% \lcmt{Remind here that we use absolute value of polack for synthetic data and not for ears ? No}

\subsection{Computational performance}

Our proposed training framework represents an additional cost at training.
Indeed, in order to train the dereverberation model,
acoustic parameters have to be computed using the acoustic analyzer,
an RIR has to be synthesized from these acoustic parameters, and the reverberation-matching loss has to be computed.
In practice, this represents an additional 502 MMACs for the forward pass of our training framework, while the original forward pass of the BiLSTM consists of 713 MMACs for instance.
% \begin{enumerate}
%     \item Computing Eq. (6): $O(FF' N \log{N})$: Around 154 MFlops.
%     \item Computing Eq. (5) $O(F F' T \log T)$
%     \item 
% \end{enumerate}
% The complexity  $O(FF' N \log{N})$
Note that at inference, the acoustic analyzer, RIR sampler, and reverberation matching loss operators are discarded, making the model performance, number of parameters and real-time compatibility equivalent to the baseline training procedure.

\subsection{Baseline}

We consider two WPE-based baselines.
First, we use the traditional unsupervised WPE method in its STFT-domain implementation (Nara-WPE~\cite{drudeNARAWPEPythonPackage2018}), with the same parameters as those used in the original article: a delay of 3 frames, a reverberation tail length of 10 frames, 3 iterations, and STFT windows of 512 samples with \qty{75}{\percent} overlap.
In addition to this purely unsupervised baseline, we also include a DNN-WPE system following the DNN-WPE formulation of~\cite{kinoshitaNeuralNetworkBasedSpectrum2017}. 
Because this method relies on strong supervision, and to ensure a fair comparison with our proposed virtually unsupervised hybrid dereverberator, we train it in a deliberately limited data regime. 
We adapt the BiLSTM dereverberation network to the Neural-WPE framework by using the BiLSTM’s output as an estimator of the clean-speech variance required by Neural-WPE. 

\lreplace{Also, strong supervision of $\bm{S}$ baseline.}{In the first experiment, where we evaluate the performance of our proposed methods in a strong supervision setting, we consider each deep-learning-based model with its original loss as a baseline.}

\section{Results and discussion}\label{sec:results_and_discussion}

This section presents the results of our experiments to train our proposed framework for the tasks of acoustic parameter estimation and dereverberation under several supervision paradigms.

The performance is evaluated using the Scale-Invariant Signal-to-Noise ratio (SISDR) \cite{rouxSDRHalfbakedWell2019}, Extended Short-Time Objective Intelligibility (ESTOI) \cite{santosImprovedNonintrusiveIntelligibility2014}, Wide-Band Perceptual Evaluation of Speech Quality (WB-PESQ)~\cite{wb_pesq} and SRMR \cite{falkNonIntrusiveQualityIntelligibility2010}  metrics.

\subsection{Dereverberation with strong supervision}

\begin{figure}
    \centering
    \includegraphics[width=\linewidth]{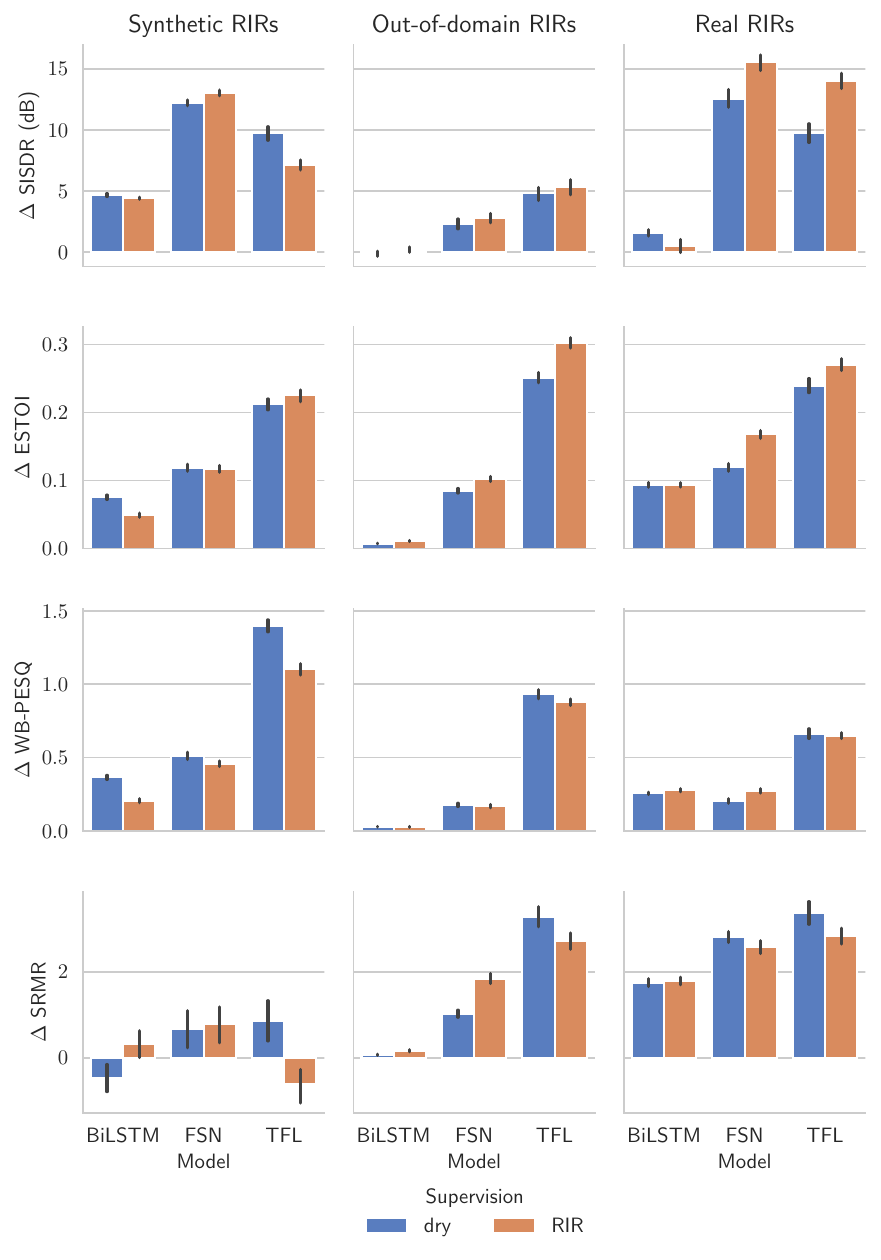}
    \caption{Dereverberation with strong supervision: 
    Comparison of the proposed training loss (supervision by the RIR) and the baseline training loss (supervision by the dry signal).
    Results are presented as the relative improvement compared to the reverberant input.
    % Relative improvement of training using ground-truth RIR compared to training using the dry signal.
    The \qty{95}{\percent} confidence intervals are indicated by black lines.}
    \label{fig:results_strong}
\end{figure}

%Figure~\ref{fig:results_strong} reports the relative improvement of training using strong supervision of the reverberation matching loss with the ground-truth RIR (defined in Eq.~\eqref{eq:strong_rir_supervision} over using the baseline supervision of dry signal.
Figure~\ref{fig:results_strong} reports the performance of the strong supervision implemented as our proposed reverberation matching loss with the ground-truth RIR (defined in Eq.~\eqref{eq:strong_rir_supervision}).
For reference, the supervision by the dry signal is also provided as a baseline.
Specifically, the baseline training losses are as follows: for the BiLSTM model, mean squared error (MSE) between the ground-truth and estimated magnitudes; for FullSubNet, MSE between the ground-truth ideal and estimated complex ratio mask; and for the TF-Locoformer, a combination of time-domain loss and a multiresolution spectral loss between the ground-truth and estimated dry signals.

On synthetic RIRs, training with supervision from the ground-truth dry signal generally outperforms supervision via the proposed reverberation matching loss. The only exceptions occur with FullSubNet, which yields higher performance on the SISDR metric when trained with the reverberation matching loss, and with TF-Locoformer, which performs better on the ESTOI metric under the same conditions.
In contrast, on real RIRs, an opposite trend is observed\lreplace{though the results are less significant.}{.} 
For all metrics except SRMR, models trained with supervision from the exact RIR either achieve better performance or exhibit differences that are not statistically significant, according to a Wilcoxon nonparametric test (p-value $< 0.001$). One explanation for this behavior is that supervision with the reverberation matching loss offers a more balanced optimization of magnitude and phase information, leading to faster convergence\lreplace{}{, which is crucial on this harder dataset}.

\subsection{Dereverberation with weak supervision}

\begin{figure}
    \centering
    \includegraphics[width=\linewidth]{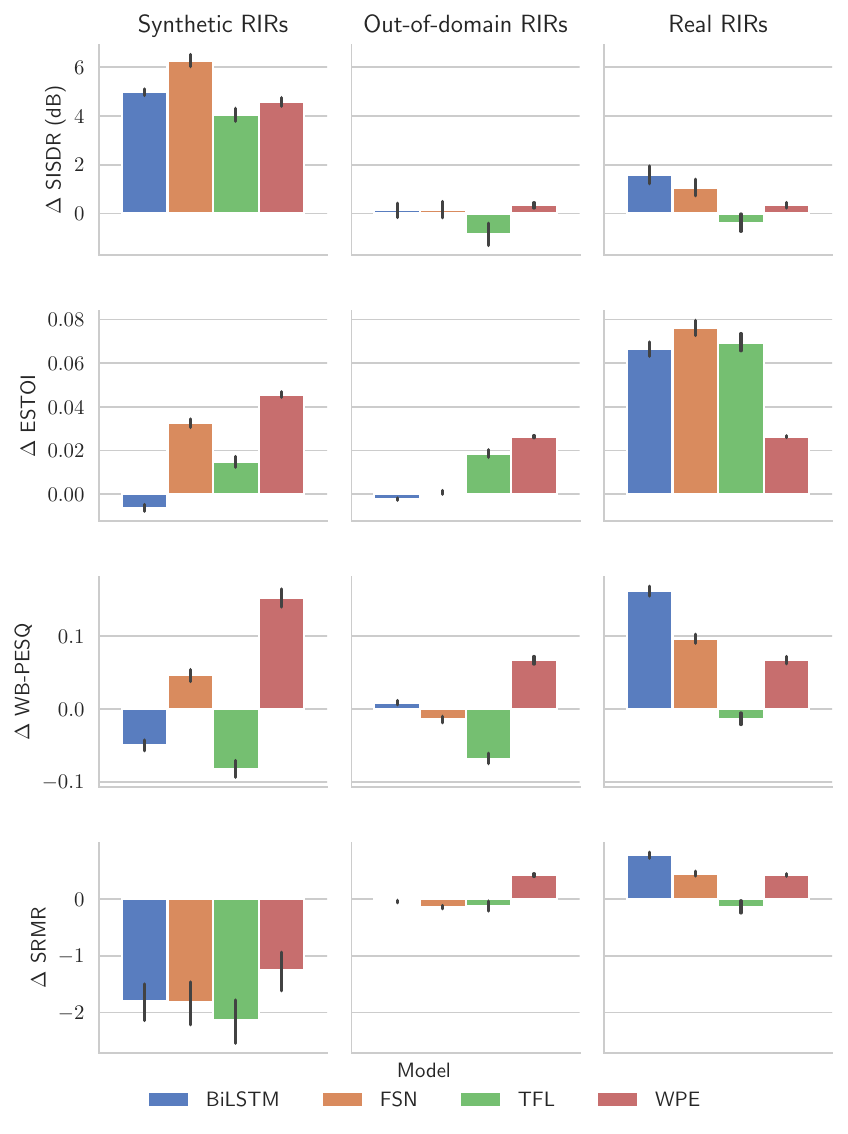}
    \caption{Weak dereverberation: 
    % Improvement of our proposed approach over the WPE baseline. 
    % Improvement over the reverberant input for all datasets.
    Comparison of the weakly-supervised models with the WPE method.
    All models (BiLSTM, FSN, TFL) are trained with the "Single" reverberation matching loss variant. 
    Results are presented as the relative improvement compared to the reverberant input.
    % Positive results indicate that our approach outperforms WPE. 
    The \qty{95}{\percent} confidence intervals are indicated by black lines.}
    \label{fig:weak_models_comparison_over_wpe}
\end{figure}

Figure~\ref{fig:weak_models_comparison_over_wpe} summarizes the performance of dereverberation models trained under weak supervision with the single variant.
The WPE baseline is provided for comparison.
% Results are first presented as the improvement of our proposed reverberation matching loss and "single" variant, compared to the WPE baseline.
% Results are first presented as the comparison of our proposed weakly supervised models trained using the reverberation matching loss and "single" variant, compared to the WPE baseline.
In comparison to the WPE baseline, few models achieve significant improvements. On real RIRs, only the BiLSTM model consistently outperforms WPE across all metrics, while FullSubNet performs better on all metrics except SRMR.
When comparing models, FullSubNet achieves equal or superior performance to BiLSTM on synthetic RIRs, while BiLSTM outperforms FullSubNet on real RIRs for SISDR, WB-PESQ, and SRMR, with FullSubNet performing better only on ESTOI. The TF-Locoformer almost consistently underperforms under the weak supervision regime, likely due to its high model capacity relative to the simplicity of the reverberation model.

\begin{figure}
    \centering
    \includegraphics[width=\linewidth]{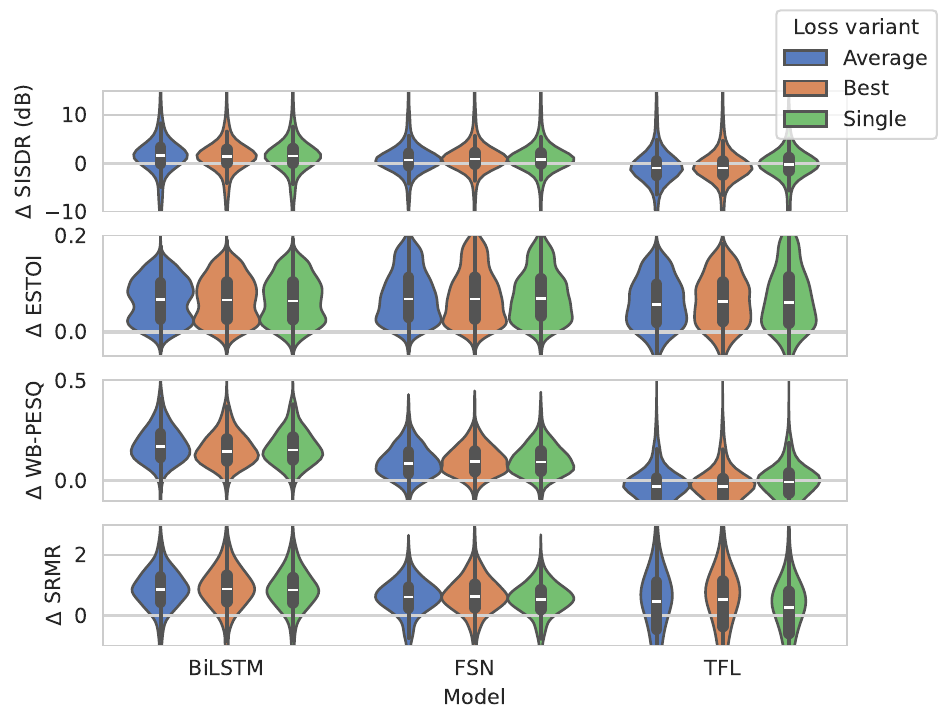}
    \caption{Weak dereverberation: Comparison of the RM loss variants on the EARS-Reverb dataset. Results are presented as the relative improvement compared to the reverberant input.}
    \label{fig:weak_variants_comparison_on_ears_reverb}
\end{figure}

Figure~\ref{fig:weak_variants_comparison_on_ears_reverb} compares the improvement of each loss variant over the reverberant input, for each source model on the EARS-Reverb dataset (since it is only single domain where our methods outperform WPE).
Across all models, datasets, and metrics, no supervision variant ("Single", "Average", or "Best" microphone strategy) consistently outperforms the others.
In most cases, the differences are not statistically significant. 
% The limited effectiveness of the "best microphone" strategy may be attributed to an insufficiently constrained training objective.
This argument, alongside with 
a deeper analysis of our
proposed reverberation sampler~\cite{marius_icassp_submission},
shows that the GradNorm approximation is not detrimental to
dereverberation performance.
For this reason, throughout the remainder of our experiments, we adopt the \textit{single} reverberation matching loss variant since it represents the lowest computational cost compared to other variants, at no cost in performance.

\begin{figure}
    \centering
    \includegraphics[width=.5\linewidth]{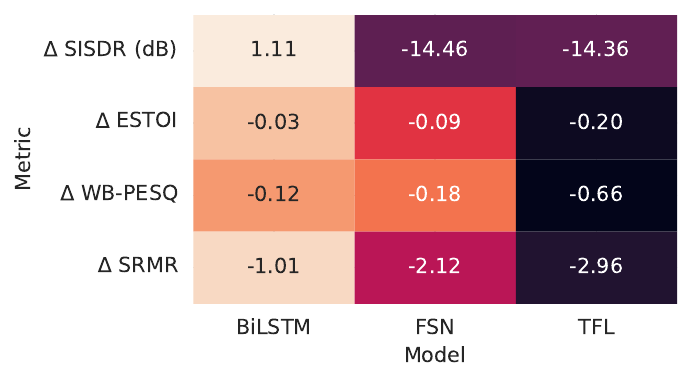}
    \caption{Weak dereverberation: Degradation caused by training using weak supervision compared to strong supervision on the EARS-Reverb dataset, using the "Single" loss variant.
    % \mfcmt{augmente la taille d'écriture c'est beaucoup trop petit ...}
    }
    \label{fig:comparison_weak_strong}
\end{figure}

Figure~\ref{fig:comparison_weak_strong} shows the degradation of using our proposed reverberation matching loss in a weak supervision setting using the oracle reverberation parameters, compared to the strong supervision of the exact RIR (from the EARS-Reverb dataset).
The results vary greatly from model to model, and remain consistent between metrics: BiLSTM shows less degradation when going from strong to weak supervision than FSN, which in turns outperforms TFL.
% The surprisingly good performance of BiLSTM compared to other models can be explained by the fact that this model is optimized for spectral magnitude masking and not for phase estimation.
% Consequently, reverberation by Polack’s model, which greatly perturbates the phase of the reverberant signal, better matches the underlying phase-agnostic assumption of the BiLSTM model than reverberation using ground-truth RIRs.
The improved performance of BiLSTM compared to other models can be explained by its relative simple architecture. BiLSTM is better suited to the approximate reverberation model used in weak supervision, and not powerful enough to optimize for the exact RIR used in strong supervision.
This underlies the need to have consistency between the deep-learning-based speech model and reverberation sampler underlying priors.

\subsection{Training the reverberation model}

\begin{figure}
    \centering
    \includegraphics[width=\linewidth]{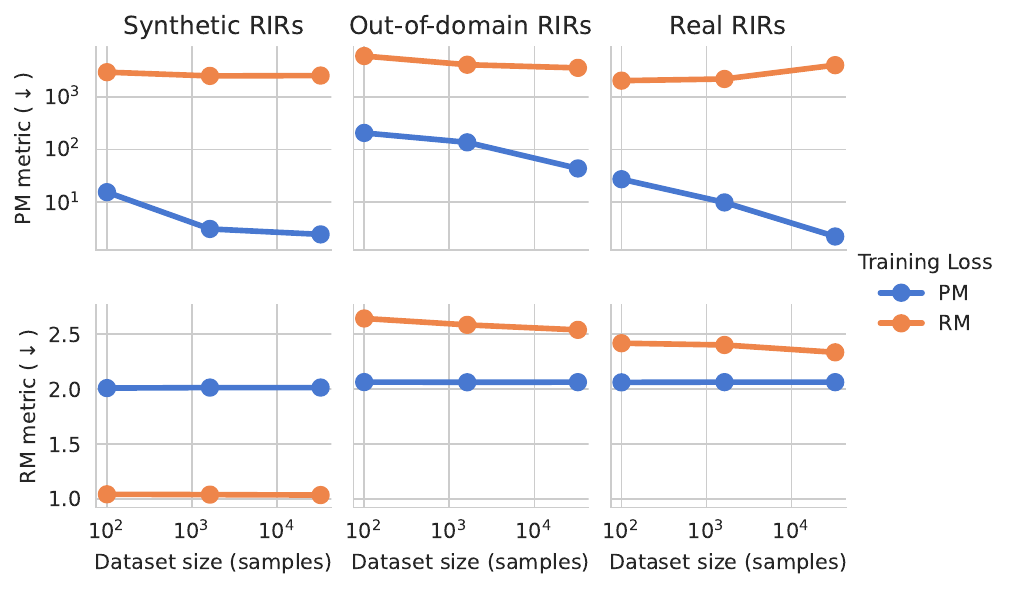}
    \caption{Acoustic parameter estimation: Comparison of the Parameter matching (PM) and our proposed Reverberation Matching (RM) loss for $\DRR$ estimation for various training set sizes.}
    \label{fig:reverb_model_results}
\end{figure}

Figure~\ref{fig:reverb_model_results} presents the performance of the acoustic analyzer $\mathcal{A}_w$ for $\DRR$ estimation under two supervision strategies: (i) the proposed reverberation matching loss, which leverages paired reverberant and dry signals, and (ii) the parameter matching loss, which requires DRR annotations for each reverberant signal.

The evaluation considers three training data regimes (full dataset, \qty{5}{\percent} subset, and 100 training samples), across synthetic, out-of-domain, and real RIRs. Performance is reported using both the PM loss (i.e., MSE between estimated and ground-truth $\DRR$) and the RM loss as evaluation metrics, to analyze cross-objective consistency.

Models trained to minimize the PM loss perform well when sufficient data is available, but their accuracy degrades significantly as training data is reduced. 
In contrast, RM loss optimization is more challenging on real RIRs, though in-domain training still outperforms training on mismatched (out-of-domain) data.
Notably, on real RIRs, models trained with RM loss increasingly diverge from their PM-trained counterparts as more data is provided. 
This suggests an inconsistency between the two objectives, likely due to late-reverberation noise in real RIRs, which affects the resynthesized target in RM but is not modeled by the RIR sampler. 
The resulting mismatch leads the model to overestimate reverberant energy, thereby distorting the inferred DRR. 
However, on synthetic RIRs, 
both models trained with PM and RM objectives yield similar results on synthetic RIRs regardless of data size.

\begin{figure}
    \centering
    \includegraphics[width=.9\linewidth]{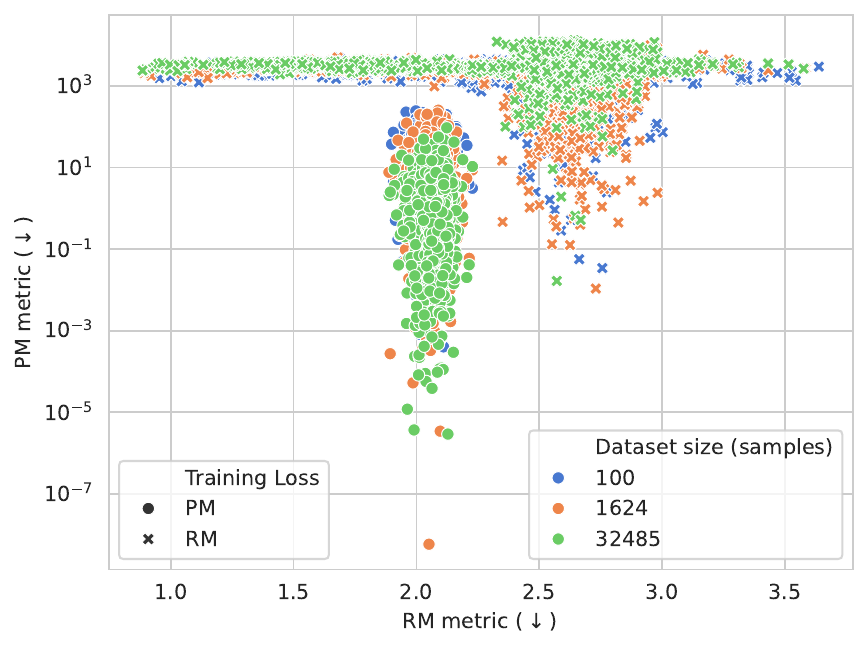}
    \caption{Acoustic parameter estimation: relative performance of each estimated blind $\DRR$ estimation sample from  EARS-Reverb on the PM and RM metrics
    % \mfcmt{tu ne peux pas spliter ta légende en deux cadres ? un à gauche avec Dataset size et l'autre à droite avec Training Loss}
    }
    \label{fig:scatterplot_pm_rm}
\end{figure}

Figure \ref{fig:scatterplot_pm_rm} provides a deeper analysis of the relationship between PM and RM metrics on the EARS-Reverb dataset.
Interestingly, RM loss performance appears largely independent of DRR estimation accuracy. Even models that overfit under PM supervision (e.g., when trained on small datasets) exhibit comparable RM performance. This indicates that optimizing the RM objective does not require precise DRR estimation and suggests that $\mathcal{A}_w$ need not be finely tuned to be effective. Finally, we observe that $\mathcal{A}_w$ achieves strong performance even with limited data, supporting our design choice to pre-train this module independently before dereverberation model training.

\subsection{Unsupervised dereverberation guided by a reverberation model}

\begin{figure}
    \centering
    \includegraphics[width=\linewidth]{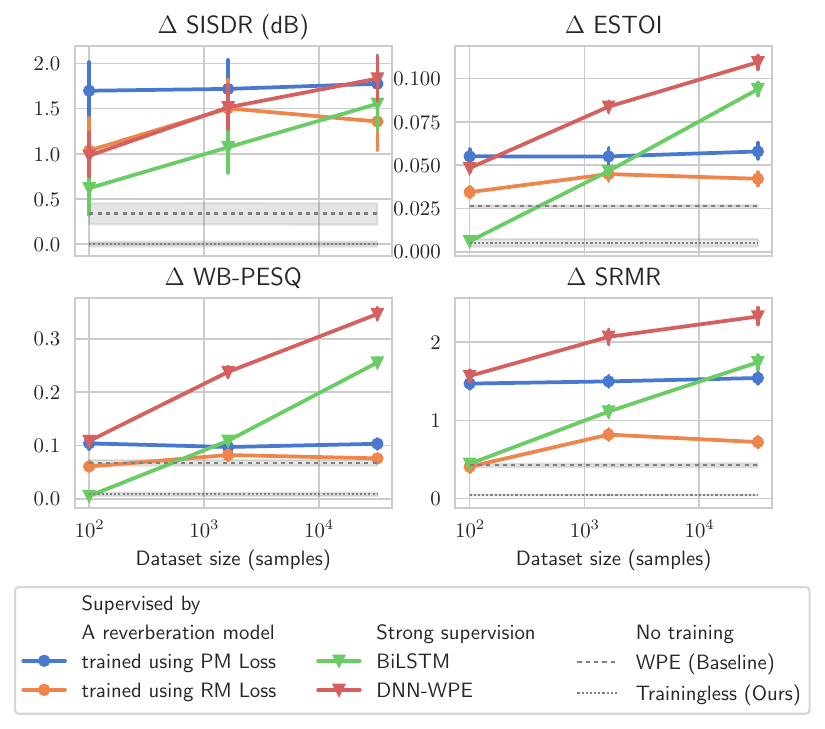}
    \caption{Unsupervised dereverberation: improvement over the reverberant input for different annotation dataset sizes. For the unsupervised methods, the x-axis represents the quantity of data used to pre-train the acoustic analyzer before training the dereverberation model. For strong supervision variants, it represents the quantity of dry data used to train the dereverberation model from scratch}
    \label{fig:results_unsupervised}
\end{figure}

In this experiment, we evaluate the impact of the blind acoustic analyzer module on the final performance of the dereverberation model trained using our proposed framework.
While the dereverberation network is retrained from scratch, the reverberation analyzers are reused from the previous experiment and remain frozen.
This allows us to examine how the dereverberation performance varies as a function of both the loss used to train the reverberation model and the quantity of supervision available during its training.
The dereverberation model itself is always trained on the full set of reverberant-only data, which reflects realistic deployment scenarios where access to clean signals or acoustic parameters is limited but not reverberant signals.
We focus on the BiLSTM model trained on real RIRs with the "single" dereverberation loss variant, as it achieved on most metrics the best performance among models that significantly outperformed the WPE baseline.
Results are presented in Figure~\ref{fig:results_unsupervised}.
The best results are obtained when the reverberation analyzer was trained using the PM loss, consistent with prior findings where PM minimized both PM and RM metrics during DRR estimation.
As expected, dereverberation performance improves with increased training data for the reverberation model.
% A notable exception is observed when using a reverberation model trained with PM loss on a large dataset, which underperforms on the SISDR and WB-PESQ metrics due to a very small degradation of the performance of the acoustic analyzer on the RM metric when the dataset size increases (as seen in the previous experiment).
% Despite this, w
We observe strong dereverberation performance when the analyzer is trained using the PM loss, even when using only 100 samples of acoustic parameters, demonstrating the robustness of our proposed framework in low-resource conditions, and outperforming the unsupervised baseline of WPE on all metrics.
Moreover, the training-less variant fails to produce any significant improvement across metrics and datasets, indicating that training the dereverberation model is necessary for effective unsupervised dereverberation.
This means that the mapping from reverberant to dry cannot be optimized on a per-sample basis, but rather at the scale of a whole dataset.

For comparison, we also evaluate strongly supervised variants of our proposed method in data-limited scenarios.
% When trained with only 100 paired dry and reverberant samples, the BiLSTM dereverberation model exhibits markedly poor performance, failing to generalize effectively.
Unless trained with the full dataset, the BiLSTM dereverberation model exhibits markedly poor performance, failing to generalize effectively.
Utilizing the same strongly-supervised BiLSTM dereverberation model in a DNN-WPE framework allows to improve the performance of this baseline, but, in the lowest-resource scenario, fails to significantly outperform our proposed unsupervised dereverberation method on all metrics except WB-PESQ.

These findings suggest that leveraging a small number of in-domain annotations to pre-train the acoustic analyzer, followed by unsupervised training of the dereverberation model, yields superior performance compared to directly training a strongly supervised dereverberation model under data-limited conditions.

In conclusion, our most efficient approach consists in pre-training an acoustic analyzer using the parameter matching loss on a dataset of 100 pairs of reverberant signals and acoustic parameters data, and use this frozen acoustic analyzer to train a dereverberation model using our proposed reverberation matching loss.

\section{Limitations}

% A limitation of our proposed UDREAM method resides in 
% the need to pre-train the acoustic analyzer using pairs of reverberant signals and acoustic parameters such as the $\rtsixty$ and $\DRR$. 
% While our methods already show competitive performance using only 100 pairs of such signals, obtaining such information limits the application of our method
% to new datasets where acoustic parameter annotations are not provided.
A first limitation of the proposed U-DREAM framework is that it requires the acoustic analyzer to be pre-trained using paired reverberant signals and corresponding acoustic parameters, such as $\rtsixty$ and $\DRR$. Although our experiments demonstrate competitive performance with as few as 100 such annotated samples, the availability of acoustic parameter labels remains a constraint and may limit applicability to datasets for which such annotations are unavailable.

We also compared our methods to the discriminative and generative baselines of the URGENT 2026 Challenge~\cite{liLessMoreData2025},
which we trained using the strong supervision of 100 pairs of reverberant and dry signals randomly sampled from the EARS dataset.
These methods were shown to benefit from carefully-curated small-scale datasets.
While our best-performing configuration significantly outperforms the generative baseline across all evaluation metrics, it remains inferior to the discriminative approach in terms of SISDR, indicating that fully supervised discriminative models may still hold an advantage when high-quality paired data are available.

Another shortcoming concerns downstream ASR performance of our methods.
% Experiments show that using our proposed method as a preprocessing step of an ASR system fails to improve the performance of said ASR system compared to providing the reverberant signal as input or using the WPE method.
When used as a preprocessing stage, the proposed dereverberation method does not yield improvements over either direct reverberant input or WPE preprocessing.
This is due to the dereverberated signals being out of the training distribution of ASR models\footnote{See \url{louis-bahrman.github.io/UDREAM/asr.html} for detailed results}.

Finally, the performance gains achieved by UDREAM remain moderate.
The observed improvements nonetheless validate the potential of leveraging reverberation models in unsupervised dereverberation frameworks.
We anticipate that further gains could be achieved by employing more expressive parametric RIR models.
% We believe that stabilzing the loss as in~\cite{marius_icassp_submission} could contribute to an increase in such performance.

\section{Conclusion}\label{sec:conclusion}

This paper formulates the dereverberation problem as a maximum likelihood estimation of the dry signal and acoustic parameters.
We propose to solve this problem using a hybrid approach that is adaptable to several supervision paradigms.
Our most data-efficient method outperforms state-of-the-art unsupervised dereverberation methods by leveraging only 100 samples of acoustic parameters such as direct-to-reverberant ratio and reverberation time.
Our experiments show that, although our approach can be generalized to any dereverberation and reverberation models, their underlying priors should match.
Future work will be dedicated to making our proposed approach robust to noisy environments and time-varying RIRs.

\bibliographystyle{IEEEtran}
\bibliography{references_clean}

\end{document}